\documentclass[11pt]{article}
\usepackage[utf8]{luainputenc}
\usepackage{color}
\usepackage{amsmath}
\usepackage{amssymb}
\usepackage{graphicx}
\usepackage{esint}

\makeatletter

\usepackage{subfigure}

\makeatother

\begin{document}
\global\long\def\half{\frac{1}{2}}
 \global\long\def\inv{^{-1}}
 \global\long\def\diag{\mathrm{diag}}
 \global\long\def\rmd{\mathrm{d}}
 \global\long\def\vect#1{\mbox{\boldmath\ensuremath{#1}}}

\title{Doubled Conformal Compactification}

\author{Zhaoyong Sun and Yu Tian\\
 \textit{School of Physics, University of Chinese Academy of Sciences,}\\
 \textit{Beijing 100049, China}\\
 \textit{State Key Laboratory of Theoretical Physics, Institute of
Theoretical Physics,}\\
 \textit{Chinese Academy of Sciences, Beijing 100190, China}\\
 \texttt{Email: sunzhaoyong11@mails.ucas.ac.cn, ytian@ucas.ac.cn}}
\maketitle
\begin{abstract}
We use Weyl transformations between the Minkowski spacetime and dS/AdS
spacetime to show that one cannot well define the electrodynamics
globally on the ordinary conformal compactification of the Minkowski
spacetime (or dS/AdS spacetime), where the electromagnetic field has
a sign factor (and thus is discountinuous) at the light cone. This
problem is intuitively and clearly shown by the Penrose diagrams,
from which one may find the remedy without too much difficulty. We
use the Minkowski and dS spacetimes together to cover the compactified
space,which in fact
leads to the doubled conformal compactification.
On this doubled conformal compactification, we obtain the globally
well-defined electrodynamics.
\end{abstract}
\section{Introduction}
Conformal transformations play an interesting role in physics and
mathematics\cite{FMS}. Conventionally, there are two meanings of
the term ``conformal transformation'' in the literature. One is
what we call the conformal coordinate transformation, which is a special
kind of coordinate transformations and whose infinitesimal generators
are the conformal Killing vectors (CKVs). The other is the local rescaling
(more commonly called the Weyl transformation), which acts directly
on the metric and can be regarded as a kind of gauge transformation.
These two meanings certainly have some relationship, but it is more
appropriate to clearly distinguish them.Unless otherwise
specified, we mean in this paper by the term ``conformal transformation''
as the conformal coordinate transformation\emph{.}

It has been shown that Maxwell's equations are invariant under the
larger conformal group \cite{E. Cunningham,Bateman-1,Rosen,FN}.
Codirla and Osborn pointed out that one can easily obtain the
electromagnetic field associated with uniformly accelerated charged
particles by the conformal invariance of the Maxwell equations with
point-like charges \cite{C. Codirla and H. Osborn}. And conformal
transformations in 2-dimension also play an important role in string
theory (see e.g. \cite{J. Polchinski}). However, as what we will
introduce in the next section, conformal transformations can not be
globally defined on the Minkowski spacetime, which means that the
conformal invariance of electrodynamics can only be regarded as
local. In fact, conformal transformations can always globally act
\textcolor{black}{on a}\textcolor{blue}{{}
}\textcolor{black}{$d$-dimensional} compactification spacetime
($d>2$),\textcolor{black}{{} which is called the conformally
compactified spacetime.} This idea inspires us to compactify the Minkowski, even dS/AdS,
spacetime.\textcolor{black}{{} In fact}, electrodynamics even cannot
be well defined globally on the ordinary conformal compactification
of the Minkowski spacetime. Instead, electrodynamics globally
defined on the double covering of that conformal compactification is
possible, which has long been noticed as mentioned in Ref. \cite{C.
Codirla and H. Osborn}. We call this double covering the doubled
conformal compactification.Interestingly, Penrose had shown the doubled Penrose diagrams in
\cite{penrose}, which is closely related to the doubled conformal
compactification here. In this paper, we try to construct the
doubled conformal conpactification to make the electrodynamics
globally defined.

Considering that the zero-radius pseudo-sphere in (4+2)-dimensional
Minkowski space can always simplify the problem and give an obvious
map about conformal compactification (see for example \cite{C. Codirla and H. Osborn,arXiv:,weinberg,P. A. M. Dirac,M. S. Costa}),
we briefly give a review of conformal compactification in the viewpoint
of the projective cone in Section 2. We show the breakdown of the
Maxwell equations on the compactification of Minkowski dS/AdS spacetime
in Section 3 and give Penrose diagrams to describe the problem in
details, in which one can find out the double covering solution, in
Section 4.

\section{General Theory and the Projective Cone}

The conformal transformations form a group, named the conformal group,
under certain conditions. It is well known that the conformal group
on the $d$-dimensional Euclidean space $\mathbb{R}^{d}$ is of $(d+1)(d+2)/2$
dimensions for $d>2$ and of infinite dimensions for $d=2$. But general
conformal transformations cannot be globally defined on such a flat
space, so this well-known result has only considered the local aspects
of conformal transformations. In fact, nontrivial conformal transformations
can be globally defined on the $d$-sphere $S^{d}$ instead, where
the conformal group is of $(d+1)(d+2)/2$ dimensions whether $d>2$
or $d=2$. There exists a conformal mapping between $\mathbb{R}^{d}$
and $S^{d}$, which smoothly extends the conventional conformal transformations
on the former to the globally defined ones on the latter. For $d=2$
only a $(d+1)(d+2)/2=6$ dimensional subgroup of the infinite-dimensional
``conformal group'' on $\mathbb{R}^{2}$ can be so extended, which
can be regarded as the ``global'' conformal group on $\mathbb{R}^{2}$.
\footnote{These two distinct conformal groups for the $d=2$ case can also be identified as the ``angle-preserving'' one and the ``circle-preserving''
one, respectively. See Ref. \cite{HS}. For another elucidation of
this problem, see Ref. \cite{FMS}, Chapter 5.}

There is exactly one point on $S^{d}$ that has no image on $\mathbb{R}^{d}$
under the conformal mapping. That point, though actually not on $\mathbb{R}^{d}$,
is called the infinity point (also known as the conformal boundary
for the case that the space has a non-positive-definite signature)
of $\mathbb{R}^{d}$. The procedure that adds the infinity point to
$\mathbb{R}^{d}$, so that the conformal transformations can be globally
defined, is known as the conformal compactification of $\mathbb{R}^{d}$.
The resulting compactified space ($S^{d}$ here) is also called the
conformal compactification (of $\mathbb{R}^{d}$). $\mathbb{R}^{d}$
and $S^{d}$ are both constant curvature (or maximal symmetry) spaces.
In differential geometry it is known that only
constant curvature spaces have $(d+1)(d+2)/2$ independent CKVs ($d>2$).
In fact, all the constant curvature spaces, with any metric signatures,
have $(d+1)(d+2)/2$-dimensional ``global'' conformal groups and
the corresponding conformal compactifications for $d\ge2$.
\footnote{$S^{d}$ is its own conformal compactification.}

To see how the doubly conformal compactification arises, let us first
have a simple review of the ordinary conformal compactification of
the Minkowski spacetime. As constant curvature spaces with the same
signature, de Sitter (dS) and anti-de Sitter (AdS) space times also
have that conformal compactification, which can be all treated from
the viewpoint of the projective cone
\footnote{Also called the null cone or the Lie sphere.
}. From now on, we concentrate on the 4-dimensional case.

The projective cone $[\mathcal{N}]$ is defined as a zero-radius pseudo-sphere
$\mathcal{N}$ in a ($4+2$)-dimensional Minkowski space:
\begin{equation}
\eta_{AB}\zeta^{A}\zeta^{B}=0,\quad(\eta_{AB})=\diag(-1,1,1,1,1,-1),\label{zero-sphere}
\end{equation}
modulo the projective equivalence relation%
\footnote{In order for $[\mathcal{N}]$ to be a (4-dimensional) manifold, the
origin $(\zeta^{A})=0$ must be excluded.%
}
\begin{equation}
(\zeta^{A})\sim\lambda(\zeta^{A}),\quad\lambda\neq0.\label{projective}
\end{equation}
The equivalence class corresponding to the point $(\zeta^{A})$ is
denoted by $[\zeta^{A}]$. The whole ($4+2$)-dimensional Minkowski
space $\mathbb{R}^{6}$ modulo the projective equivalence relation
(\ref{projective}) is the 5-dimensional projective space $\mathbb{R}P^{5}$,
so $[\mathcal{N}]$ is a submanifold of $\mathbb{R}P^{5}$. It is
obvious that the pseudo-sphere (\ref{zero-sphere}) and the equivalence
relation (\ref{projective}) are both invariant under general $O(2,4)$
transformations, among which a $\mathbb{Z}_{2}$ antipodal reflection
\begin{equation}
(\zeta^{A})\rightarrow-(\zeta^{A})\label{antipodal}
\end{equation}
acts trivially on $[\mathcal{N}]$.

Then it can be shown that $[\mathcal{N}]$ is the conformal compactification
of all the Minkowski, dS and AdS space times, where the $O(2,4)/\mathbb{Z}_{2}$
transformations act as conformal transformations on these space times.
These different space times can be regarded as different choices of
representatives in the equivalence classes on $\mathcal{N}$. In fact,
these constant curvature space times correspond to choose representative
points by intersecting $\mathcal{N}$ with hyperplanes in $\mathbb{R}^{6}$,
where the metrics on them are naturally induced from $\eta_{AB}$.
For the hyperplane $\mathcal{P}_{a}$:
\begin{equation}
a_{A}\zeta^{A}=1\label{}
\end{equation}
with $(a_{A})\neq0$ the normal vector, it can be shown that the intersection
manifold is characterized by
\begin{equation}
S=\eta^{AB}a_{A}a_{B}\label{}
\end{equation}
as follows:
\begin{itemize}
\item $S<0$: dS spacetime,
\item $S=0$: Minkowski spacetime,
\item $S>0$: AdS spacetime,
\end{itemize}
where in any case $S$ is just the scalar curvature of the intersection
manifold .

More concretely, to get the Minkowski spacetime we just choose a light-like
normal vector
\begin{equation}
(a_{A})=(0,0,0,0,1,1).\label{}
\end{equation}
Thus the intersection $M$ of $\mathcal{N}$ and the corresponding
$\mathcal{P}_{a}$ is flat with respect to the metric induced from
$\eta_{AB}$, and can be parametrized by
\begin{equation}
x^{\mu}=L\frac{\zeta^{\mu}}{\zeta^{+}},\quad\mu=0,\cdots,3,\label{}
\end{equation}
with $L$ an arbitrary length scale parameter and $\zeta^{\pm}=\zeta^{5}\pm\zeta^{4}$
the lightcone coordinates. The induced metric is proportional to vect
\begin{equation}
\rmd s^{2}=\eta_{\mu\nu}\rmd x^{\mu}\rmd x^{\nu},\label{}
\end{equation}
so $x^{\mu}$ is just the Cartesian coordinates on $M$. Then it is
straightforward to show that the $O(2,4)/\mathbb{Z}_{2}$ transformations
act as conformal transformations on $M$ (\textcolor{black}{for more
details, see \cite{penrose}}). Note that some equivalence classes
on $\mathcal{N}$, which correspond to the infinity points of $M$,
have no representatives on $\mathcal{P}_{a}$. They constitute precisely
the intersection of $\mathcal{N}$ and the hyperplane
\begin{equation}
a_{A}\zeta^{A}=0,\label{}
\end{equation}
which is parallel to $\mathcal{P}_{a}$. Adding those infinity points
to $M$ produces its conformal compactification $[\mathcal{N}]$,
whose $O(2,4)/\mathbb{Z}_{2}$ action is fully well-defined.

The dS and AdS space times can be obtained by choosing typically
\begin{equation}
(a_{A})=(0,0,0,0,0,1)\label{}
\end{equation}
and
\begin{equation}
(a_{A})=(0,0,0,0,1,0),\label{}
\end{equation}
respectively. However, they also have infinity points, or conformal
boundary, to be included for a fully well-defined $O(2,4)/\mathbb{Z}_{2}$
conformal action, since no single hyperplane can contain representatives
of all the equivalence classes on $\mathcal{N}$. To remedy this problem,
one may use general hypersurfaces (of antipodal symmetry) to intersect
$\mathcal{N}$. It can be shown that these intersection manifolds
are all conformally flat,%
\footnote{So they are all related by Weyl transformations (also called conformal
mappings), at least locally.%
} and that the $O(2,4)/\mathbb{Z}_{2}$ transformations induce conformal
transformations on them. The simplest choice of this hypersurface
is a 5-sphere
\begin{equation}
\delta_{AB}\zeta^{A}\zeta^{B}=2L^{2},\label{5-sphere}
\end{equation}
which intersects all the equivalence classes on $\mathcal{N}$ precisely
twice. The intersection of $\mathcal{N}$ and the 5-sphere (\ref{5-sphere})
is an $S^{1}\times S^{3}$:
\begin{equation}
\left\{ \begin{array}{rcl}
(\zeta^{0})^{2}+(\zeta^{5})^{2} & = & L^{2},\\
(\zeta^{1})^{2}+(\zeta^{2})^{2}+(\zeta^{3})^{2}+(\zeta^{4})^{2} & = & L^{2}.
\end{array}\right.\label{intersection}
\end{equation}
Upon the antipodal identification (\ref{antipodal}), one sees that
$[\mathcal{N}]$ is of topology
\begin{equation}
S^{1}\times S^{3}/\mathbb{Z}_{2},\label{}
\end{equation}
which is actually homeomorphic to $S^{1}\times S^{3}$.

Although there is no natural metric defined on $[\mathcal{N}]$, there
is an induced metric on the intersection manifold (\ref{intersection})
(modulo the antipodal identification), which is called $N$ hereafter.
$N$ is conformally flat and has a globally defined $O(2,4)/\mathbb{Z}_{2}$
conformal group, so it can be regarded as a metrical realization of
$[\mathcal{N}]$, being the conformal compactification of all the
Minkowski, dS and AdS space times.

The above construction of conformal compactification can be extended
to any dimension and any metric signature. However, for even $d$
it can be shown that $S^{1}\times S^{d}/\mathbb{Z}_{2}$ is not homeomorphic
to $S^{1}\times S^{d}$ but is a non-orientable manifold, so the conformal
compactification of ($1+d$)-dimensional spacetime is not simple and
in some sense not suitable to be a spacetime. One immediately sees
that this shortcoming can be simply overcome by discarding the $\mathbb{Z}_{2}$
antipodal identification. That is the doubly conformal compactification,
which can be realized by replacing the projective equivalence relation
(\ref{projective}) with the pseudo-projective one:
\begin{equation}
(\zeta^{A})\sim\lambda(\zeta^{A}),\quad\lambda>0.\label{}
\end{equation}
For the $d=3$ case, we use $[\mathcal{N}]_{+}$ to denote ${\cal N}$
modulo the above equivalence relation, i.e., the doubly conformal
compactification of all the Minkowski, dS and AdS space times, whose
metrical realization is exactly the intersection manifold (\ref{intersection}),
denoted by $2N$. Correspondingly, the conformal group on (\ref{intersection})
is $O(2,4)$ instead of $O(2,4)/\mathbb{Z}_{2}$.

\section{The breakdown of the Maxwell Equation on the Compactification of
Minkowski Spacetime}

  We have shown the geometrical process of compactification.
It is the most important that physical considerations support the
introduction of the doubly conformal compactification. At this section
we will introduce the discontinuity of electrodynamics caused by compactifing
the Minkowski spacetime. The action functional of electrodynamics
in general space times is
\begin{equation}
S_{\mathrm{EM}}=\int d^{4}x\sqrt{-g}\left(-\frac{1}{4}g^{\mu\alpha}g^{\nu\beta}F_{\mu\nu}F_{\alpha\beta}+g^{\mu\nu}J_{\mu}A_{\nu}\right),\quad F_{\mu\nu}=\partial_{\mu}A_{\nu}-\partial_{\nu}A_{\mu}.\label{action}
\end{equation}
In this paper the metrics always have a $(-,+,+,+)$ signature. For
the Minkowski spacetime with Cartesian coordinates ($g_{\mu\nu}=\eta_{\mu\nu}$),
the above action functional is invariant (up to a boundary term) under
the conformal transformations provided that $A_{\mu}$ and $J_{\mu}$
transform as
\begin{equation}
A_{\mu}(x)=\frac{\partial\tilde{x}^{\nu}}{\partial x^{\mu}}\tilde{A}_{\nu}(\tilde{x}),\quad J_{\mu}(x)=\Omega^{2}\frac{\partial\tilde{x}^{\nu}}{\partial x^{\mu}}\tilde{J}_{\nu}(\tilde{x}),\label{}
\end{equation}
respectively, where the transformation of $A_{\mu}$ is just trivial
and that of $J_{\mu}$ contains a conformal factor $\Omega^{2}$ defined
by
\begin{equation}
\frac{\partial\tilde{x}^{\mu}}{\partial x^{\rho}}\frac{\partial\tilde{x}^{\nu}}{\partial x^{\sigma}}\eta_{\mu\nu}=\Omega^{2}\eta_{\rho\sigma}.\label{}
\end{equation}

In fact, the action functional (\ref{action}) in general space times
is invariant under, in addition to the diffeomorphism, the Weyl transformation
\begin{equation}
g_{\mu\nu}(x)=k^{-2}(x)g'_{\mu\nu}(x),\label{}
\end{equation}
provided that $A_{\mu}$ (and thus $F_{\mu\nu}$) is invariant and
$J_{\mu}$ transforms as
\begin{equation}
J_{\mu}(x)=k^{2}(x)J'_{\mu}(x).\label{J_Weyl}
\end{equation}
Since the dS, AdS and $N$ space times can all be obtained by Weyl
transformations from the Minkowski spacetime, where the Cartesian
coordinates $x^{\mu}$ become the conformally flat coordinates on
these space times, we can easily map the electrodynamics from the
Minkowski spacetime to them.

The conformally flat coordinates on the dS spacetime can be obtained
by the stereographic projection \cite{Tian}, with the Weyl factor
\begin{equation}
k_{{\rm dS}}^{-2}(x)=\left(1+\frac{x^{2}}{4R^{2}}\right)^{2},\quad x^{2}=\eta_{\mu\nu}x^{\mu}x^{\nu},\label{dS_factor}
\end{equation}
with $R$ the dS radius. The AdS case is simply achieved by replacing
$R^{2}$ with $-R^{2}$ for most of the dS expressions, so we only
mention the dS case. The Weyl factor for certain conformally flat
coordinates on $N$ can be shown to be
\begin{equation}
k_{N}^{-2}(x)=1+\frac{x^{0}x^{0}+\sum_{i}x^{i}x^{i}}{2L^{2}}+\left(\frac{x^{2}}{4L^{2}}\right)^{2},\label{N_factor}
\end{equation}
which is positive definite reflecting the fact that $N$ has no conformal
boundary.

The action functional (\ref{action}) with $g_{\mu\nu}=\eta_{\mu\nu}$
\begin{equation}
S_{\mathrm{EM}}=\int d^{4}x\left(-\frac{1}{4}\eta^{\mu\alpha}\eta^{\nu\beta}F_{\mu\nu}F_{\alpha\beta}+\eta^{\mu\nu}J_{\mu}A_{\nu}\right)\label{flat action}
\end{equation}
leads to the familiar equation of motion
\begin{equation}
\eta^{\alpha\mu}\partial_{\alpha}F_{\mu\nu}=J_{\nu}.\label{EOM}
\end{equation}
The simplest solution to the above equation is
the Coulomb field for a static point charge:
\begin{equation}
E_{i}=F_{i0}=\frac{e}{4\pi}\frac{x^{i}}{|\vect x|^{3}},\quad B_{i}=\half\epsilon^{ijk}F_{jk}=0,\quad J_{0}=e\delta^{3}(\vect x),\quad J_{i}=0.\label{fundamental}
\end{equation}
General solutions to equation (\ref{EOM})
are just linear combinations of this fundamental
solution. Of special interest is the electromagnetic field
associated with uniformly accelerated point charge, which can be
obtained from the solution (\ref{fundamental}) by conformal
transformations \cite{C. Codirla and H. Osborn}. There one sees
that, however, an additional sign factor $\epsilon(\Omega)$ has to
be inserted (or equivalently, discarded) to attain a globally
defined solution. That, in fact, already indicates the doubly
conformal compactification. In the following, we will use Weyl
transformations to map the solution (\ref{fundamental}) onto the dS
and $N$ space times, where the doubly conformal compactification is
shown to be necessary. For simplicity, we take $R=1/2$ in equation
(\ref{dS_factor}) and massless $L=1/2$ in equation (\ref{N_factor}).

First we consider the dS case. In this case we have from equations
(\ref{J_Weyl}) and (\ref{dS_factor})
\begin{equation}
J'_{0}=e(1+x^{2})^{2}\delta^{3}(\vect x),\quad J'_{i}=0,\label{dS_J}
\end{equation}
while $E'_{i}$ and $B'_{i}$ are the same as $E_{i}$ and $B_{i}$ in
equation (\ref{fundamental}), respectively. Note that there are
actually two antipodal point charges here, due to the conformal
boundary $1+x^{2}=0$ separating the world line $\vect x=0$ into two
parts. Properly speaking, the line $\vect x=0$ is
separated into three segments, but the outer two of them are joined
through the Minkowski conformal infinity, as shown in figure
\ref{Penrose-CFdS}, where the dash line KN can be
regarded as the world line of the charges in dS spacetime. For
convenience, we omit the prime in these notations in the following
equations. Since one patch of conformally flat coordinates
cannot cover the whole dS spacetime,%
\footnote{The uncovered part corresponds to (part of) the conformal boundary
of the Minkowski spacetime.%
} other coordinate patches must also be used to check whether the above
solution is globally defined on the dS spacetime. It is easy to see
that one more (conformally-flat-coordinate) patch is sufficient, which
can be viewed as the inversion
\begin{equation}
x^{\mu}=\frac{\tilde{x}^{\mu}}{\tilde{x}^{2}}\label{inversion}
\end{equation}
of the original conformally flat coordinates. ( One can define
special conformal transformations by combining an inversion with a
translation and then another inversion \cite{C. Codirla and H.
Osborn}.) Under the coordinate transformation (\ref{inversion}), we
have
\begin{equation}
\tilde{J}_{\mu}(\tilde{x})=\frac{\partial x^{\nu}}{\partial\tilde{x}^{\mu}}J_{\nu}(x)=\left[\frac{\delta_{\mu}^{\nu}}{\tilde{x}^{2}}-2\frac{\tilde{x}^{\nu}\tilde{x}_{\mu}}{(\tilde{x}^{2})^{2}}\right]J_{\nu}(x),\quad\tilde{x}_{\mu}=\eta_{\mu\nu}\tilde{x}^{\nu},
\end{equation}
which means
\begin{eqnarray}
\tilde{J}_{0}(\tilde{x}) & = & \frac{J_{0}(x)}{\tilde{x}^{2}}-2\tilde{x}_{0}\frac{\tilde{x}^{0}J_{0}(x)+\tilde{x}^{i}J_{i}(x)}{(\tilde{x}^{2})^{2}}\nonumber \\
 & = & e[1+(\tilde{x}^{2})\inv]^{2}\left[\frac{1}{\tilde{x}^{2}}+\frac{2\tilde{t}^{2}}{(\tilde{x}^{2})^{2}}\right]\delta^{3}\left(\frac{\tilde{\vect x}}{\tilde{x}^{2}}\right)\nonumber \\
 & = & e(1-\tilde{t}^{2})^{2}\delta^{3}(\tilde{\vect x})\label{dS_inverse_J}
\end{eqnarray}
with $\tilde{t}=\tilde{x}^{0}$, and
\begin{equation}
\tilde{J}_{i}(\tilde{x})=\frac{J_{i}(x)}{\tilde{x}^{2}}-2\tilde{x}_{i}\frac{\tilde{x}^{0}J_{0}(x)+\tilde{x}^{i}J_{i}(x)}{(\tilde{x}^{2})^{2}}=0.\label{}
\end{equation}
At the same time, the electromagnetic field transforms as
\begin{equation}
\tilde{F}_{\mu\nu}=\left[\frac{\delta_{\mu}^{\rho}}{\tilde{x}^{2}}-2\frac{\tilde{x}^{\rho}\tilde{x}_{\mu}}{(\tilde{x}^{2})^{2}}\right]\left[\frac{\delta_{\nu}^{\sigma}}{\tilde{x}^{2}}-2\frac{\tilde{x}^{\sigma}\tilde{x}_{\nu}}{(\tilde{x}^{2})^{2}}\right]F_{\rho\sigma}=\frac{F_{\mu\nu}}{(\tilde{x}^{2})^{2}}-2\frac{\tilde{x}^{\rho}\tilde{x}_{\mu}}{(\tilde{x}^{2})^{3}}F_{\rho\nu}-2\frac{\tilde{x}^{\sigma}\tilde{x}_{\nu}}{(\tilde{x}^{2})^{3}}F_{\mu\sigma},\label{}
\end{equation}
which means
\begin{eqnarray}
\tilde{F}_{i0} & = & (\tilde{x}^{2})^{-3}[\tilde{x}^{2}F_{i0}-2\tilde{x}_{i}\tilde{x}^{j}F_{j0}+2\tilde{t}(\tilde{t}F_{i0}+\tilde{x}^{j}F_{ij})]\nonumber \\
 & = & \frac{e}{4\pi}(\tilde{x}^{2})^{-3}\left[(\tilde{t}^{2}+\tilde{\vect x}^{2})\frac{\tilde{x}^{i}/\tilde{x}^{2}}{|\tilde{\vect x}/\tilde{x}^{2}|^{3}}-2\tilde{x}^{i}\frac{\tilde{\vect x}^{2}/\tilde{x}^{2}}{|\tilde{\vect x}/\tilde{x}^{2}|^{3}}\right]\nonumber \\
 & = & -\epsilon(\tilde{x}^{2})\frac{e}{4\pi}\frac{\tilde{x}^{i}}{\tilde{\vect x}^{2}}\label{tilde_E}
\end{eqnarray}
with $\epsilon(\tilde{x}^{2})$ the sign of $\tilde{x}^{2}$ and
\begin{equation}
\tilde{F}_{ij}=-2(\tilde{x}^{2})^{-3}\tilde{t}(\tilde{x}_{i}F_{0j}+\tilde{x}_{j}F_{i0})=0.\label{tilde_B}
\end{equation}

The discontinuity of $\tilde{F}_{i0}$ from the $\epsilon(\tilde{x}^{2})$
factor in equation (\ref{tilde_E}) indicates the breakdown of the
Maxwell equation at $\tilde{x}^{2}=0$. In fact, one may roughly understand
this breakdown by the usual argument that there can be no net charge
on compact spaces, since the dS spacetime can be viewed as an expanding
$S^{3}$ and it can be shown from equations (\ref{dS_J}) and (\ref{dS_inverse_J})
that there are two charges of the same signature on any one of the
$S^{3}$ simultaneity hypersurfaces.

Also, it's interesting to consider about Maxwell equation with the
magnetic and electric charges. The solution (\ref{fundamental}) is
changed to be:

\[
E_{i}=F_{i0}=\frac{e}{4\pi}\frac{x^{i}}{\left|\vect x\right|^{3}},\quad B_{i}=\frac{1}{2}\epsilon^{ijk}F_{jk}=\frac{e_{m}}{4\pi}\frac{x^{i}}{\left|\vect x\right|^{3}},\quad J=e\delta^{3}\left(\vect x\right),\quad J_{0}^{m}=e_{m}\delta^{3}\left(\vect x\right)
\]

where the $e_{m}$ and $J^{m}$ are the magnetic charge and current.
With the second equation, one can easily see that the field

\[
F_{ij}=\epsilon_{kij}B_{k}=\frac{e_{m}}{4\pi}\frac{\epsilon_{kij}x^{k}}{\left|\vect x\right|^{3}}\neq0
\]

So the transformation of the field seems different
with(\ref{tilde_E}) and (\ref{tilde_B}). In fact, after adding the
magnetic charge, there will be a none zero term in $\tilde{F_{i0}}$,
which means that the magnetic charges will contribute to the electric
field in the conformal transformation. But the discontinuity of the
$\tilde{F_{i0}}$ at $\vect x=0$ still exists since the discontinuity
is \textcolor{black}{actually caused by the $\left|\vect x\right|^{3}$
term. Evidently, the magnetic field is not zero, but a function containing
the $\left|\vect x\right|^{3}$ term also has a $\varepsilon\left(\tilde{x}^{2}\right)$
factor.} This indicates the discontinuity of the magnetic field at
$\tilde{x}^{2}=0$. One can roughly understand this by thinking about
the symmetry of electric and the magnetic charges.

\section{Penrose Diagrams and the Doubled Covering}

\begin{figure}[htbp]
\centering\subfigure[Penrose diagram of the dS spacetime in
conformally flat coordinates, with identification $KL=MN$ and
$KM=LN$, which can be compared with figure 1 in \cite{Tian}. The
solid line segments $GH$ and $IJ$ are its conformal boundary. The
dashed line segment stands for the world line(s) of the point
charge(s).]{\label{Penrose-CFdS}%

\begin{minipage}[c]{0.8\textwidth}%
\centering\includegraphics[scale=0.9]{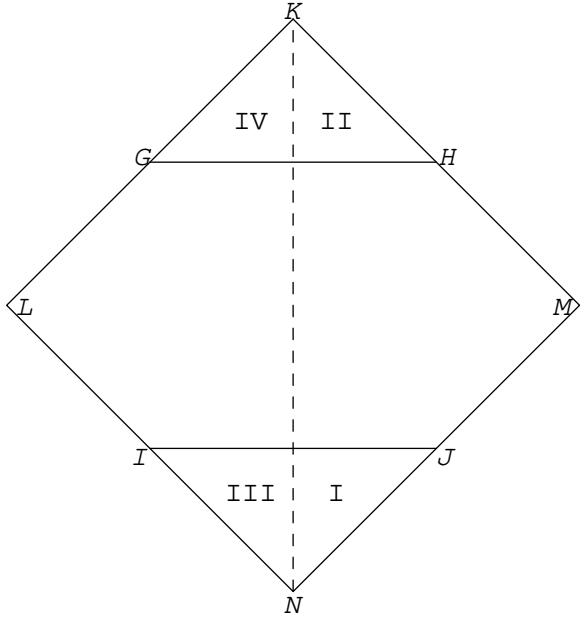} %
\end{minipage}}\vspace{0.4cm}
 \subfigure[Inversion (\ref{inversion}) of figure
\ref{Penrose-CFdS}. The points $K$, $L$, $M$ and $N$ are all
transformed to the origin, and the triangles (numbered ``I" to
``IV") in figure \ref{Penrose-CFdS} to the corresponding positions
in this figure, respectively.]{\label{Penrose-InvdS}%
\begin{minipage}[c]{0.8\textwidth}%
\centering\includegraphics[scale=0.9]{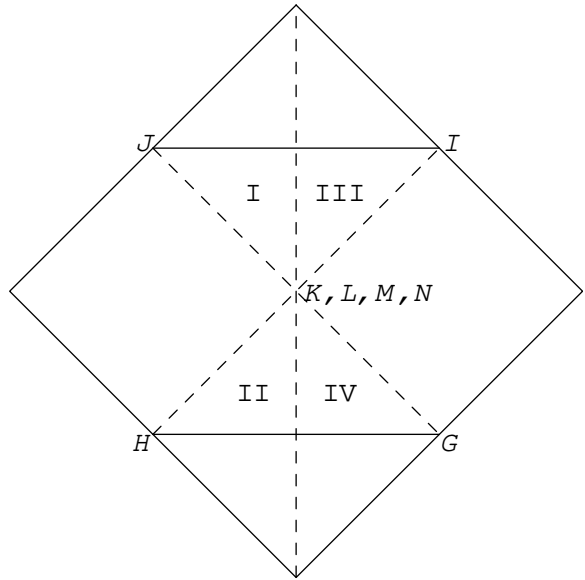} %
\end{minipage}}\caption{An illustration of the point charge(s) in the dS spacetime with Penrose
diagrams.}

\label{Penrose}
\end{figure}

The above discussion can be illustrated with Penrose diagrams, as
in figure \ref{Penrose}. The lightcone $\tilde{x}^{2}=0$ corresponds
to the dashed line segments $JG$ and $IH$ in figure \ref{Penrose-InvdS}.
Then it is easy to see that the discontinuity of $\tilde{F}_{i0}$
at $\tilde{x}^{2}=0$ can be removed by reversing the sign of the
electromagnetic field and electric current:
\begin{equation}
A_{\mu}\rightarrow-A_{\mu},\quad J_{\mu}\rightarrow-J_{\mu},\label{reversion}
\end{equation}
in the regions ``I'' to ``IV'', so that the Maxwell equation
\begin{equation}
\eta^{\alpha\mu}\partial_{\alpha}F_{\mu\nu}=k_{{\rm dS}}^{2}(x)J_{\nu}\label{}
\end{equation}
can be satisfied on the whole dS spacetime. We explicitly write this
global solution as
\begin{equation}
E_{i}=\epsilon(1+x^{2})\frac{e}{4\pi}\frac{x^{i}}{|\vect x|^{3}},\quad B_{i}=0,\quad J_{0}=\epsilon(1+x^{2})e(1+x^{2})^{2}\delta^{3}(\vect x),\quad J_{i}=0.\label{global}
\end{equation}
For this solution the two antipodal point charges on the dS spacetime
have opposite signs.%
\footnote{It is interesting if this fact has something to do with the arguments
given in \cite{PSV}, which is from a completely different point of
view.%
}

\begin{figure}[htbp]
\centering \includegraphics[scale=0.9]{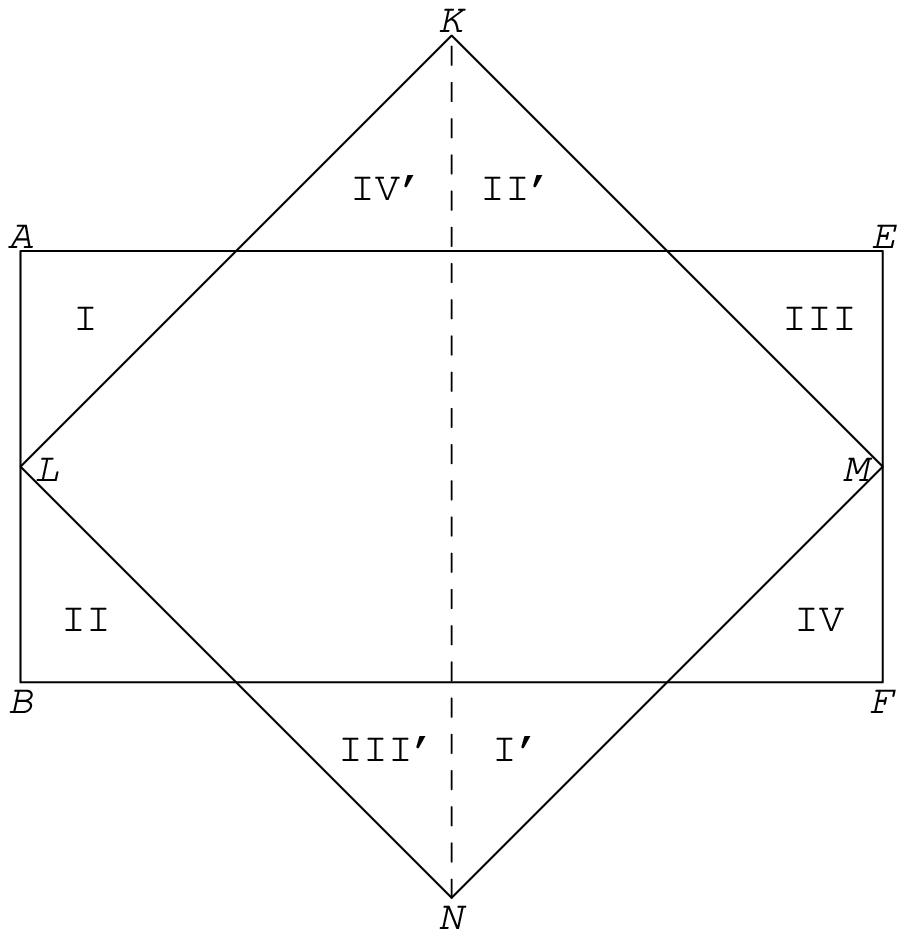}
\caption{Superposition of the familiar Penrose diagrams of dS and
Minkowski space times, where extension of the ordinary conformal
compactification arises.}

\label{Penrose-MdS}
\end{figure}

The Penrose diagram of dS spacetime as figure \ref{Penrose-CFdS}
is not the familiar one, but has the shape of that of the Minkowski
spacetime. It turns out to be interesting that we superpose the familiar
Penrose diagrams of dS and Minkowski space times, as in figure \ref{Penrose-MdS}.
There the cylinder $AEFB$ (with identification $AB$=$EF$) is the
Penrose diagram of dS spacetime, while the diamond $KLNM$ is that
of Minkowski spacetime. If we identify the regions ``I'' to ``IV''
with ``I$'$'' to ``IV$'$'', respectively, we obtain the ordinary
conformal compactification of dS and Minkowski space times (see also
\cite{Tian}). However, we have seen that for globally defined solutions
on dS and Minkowski space times (\ref{fundamental}) and (\ref{global})
the electromagnetic field and electric current in the regions ``I''
to ``IV'' differ from that in ``I$'$'' to ``IV$'$'' by a sign,
up to positive Weyl factors. In order to find a conformal compactification
of the Minkowski and dS/AdS space times where the electrodynamics
can be globally defined, then, instead of identifying the regions
``I'' to ``IV'' with ``I$'$'' to ``IV$'$'' immediately,
one should take them as, actually, antipodal regions%
\footnote{This ``antipodal'' refers to the 6-dimensional one (\ref{antipodal}),
as can be seen more clearly in the following discussion.%
} on the doubly conformal compactification of these space times. In
other words, we use Minkowski spacetime and dS spacetime to cover
different parts of this doubly conformal compactification.

\begin{figure}[htbp]
\centering \includegraphics[scale=0.9]{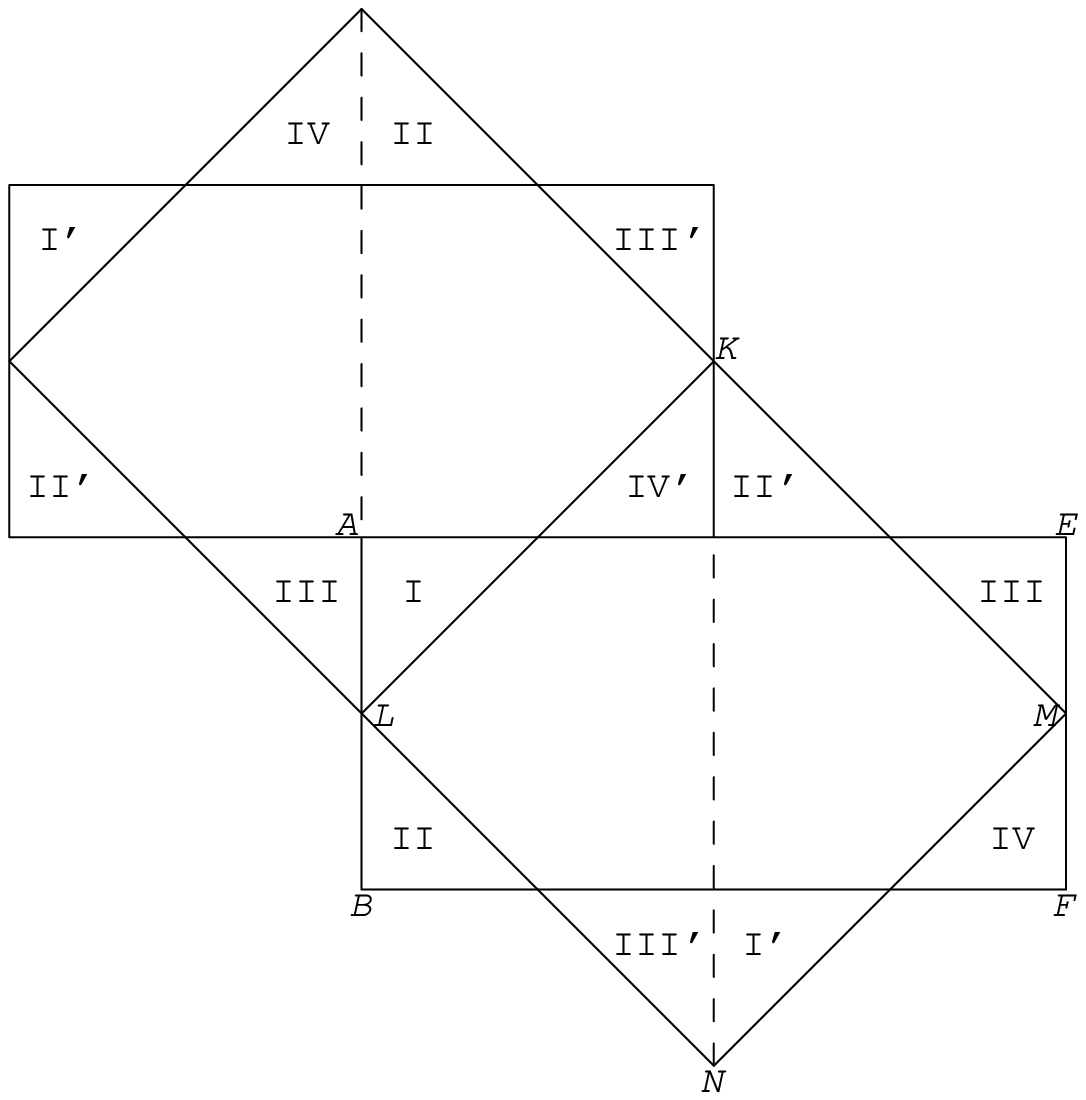}
\caption{Double extension of figure \ref{Penrose-MdS}.}

\label{Penrose-Ext}
\end{figure}

\begin{figure}[htbp]
\centering\subfigure[Double extension of the Penrose diagram of dS
spacetime, with the usual cylindrical identification $B'B=F'F$,
which is conformally compactified by identifying $B'F'=BF$.]{\label{Penrose-DdS}%
\begin{minipage}[c]{0.8\textwidth}%
\centering\includegraphics[scale=0.9]{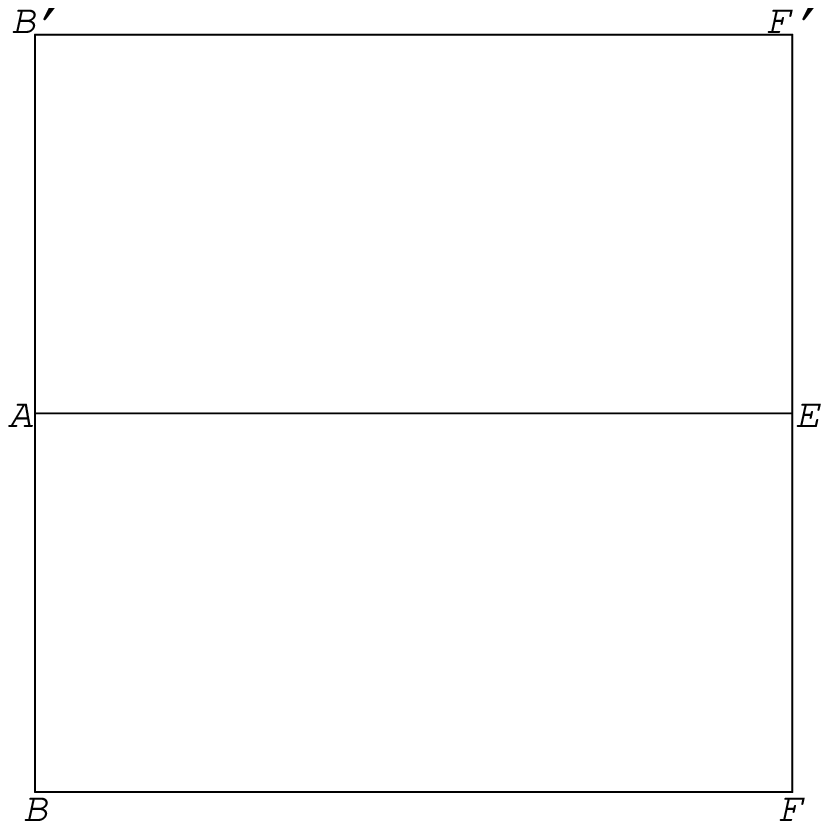} %
\end{minipage}}\vspace{0.4cm}
 \subfigure[Double extension of the Penrose diagram of Minkowski
spacetime, which is conformally compactified by identifying
$OP=QR$ and $OQ=PR$.]{\label{Penrose-DMink}%
\begin{minipage}[c]{0.8\textwidth}%
\centering\includegraphics[scale=0.9]{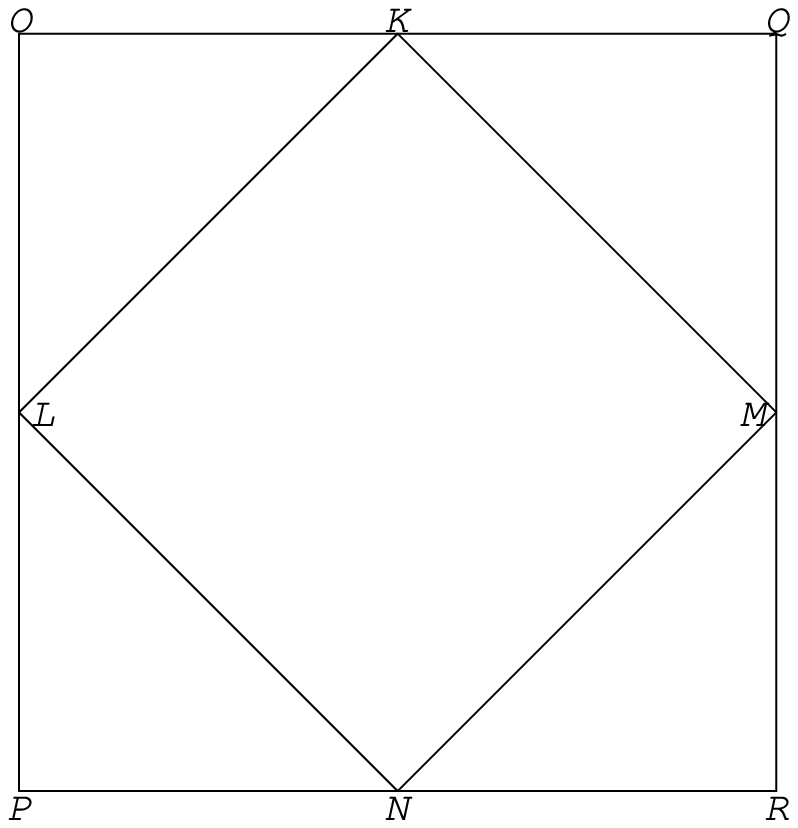} %
\end{minipage}}\caption{An illustration of the doubly conformal compactification of dS and
Minkowski space times with Penrose diagrams.}

\label{Penrose-Double}
\end{figure}

Although we have not seen in figure\ref{Penrose-MdS}
the whole of the doubly conformal compactification of Minkowski and
dS/AdS space times, it is rather straightforward to construct it based
on the above analysis. From figure \ref{Penrose-MdS} it is clear
that antipodal points have relative coordinates $(\pm1,\pm1)$ on
the Penrose diagram, where we have taken $AB$ as the length unit.
The antipode of antipode, with relative coordinate $(\pm2,0)$ or
$(0,\pm2)$, should be itself. In fact, the electromagnetic field
and electric current are of the same value, up to positive Weyl factors,
at these points, so it is safe to identify these points in the conformal
sense. A thus extended version of figure \ref{Penrose-MdS} is figure
\ref{Penrose-Ext}. The doubly conformal compactification of dS and
Minkowski space times is shown, more clearly, in figure \ref{Penrose-Double}.
Note also that any pair of points related by an inversion-like transformation
\begin{equation}
x^{\mu}\rightarrow-\eta_{\mu\nu}\frac{x^{\nu}}{x^{2}}=-\frac{x_{\mu}}{x^{2}}\label{inversion-like}
\end{equation}
can be viewed as to have relative coordinates $(\pm1,0)$ or $(0,\pm1)$
on these diagrams.

Then we consider the $N$ case. Similarly, we have from equations
(\ref{J_Weyl}) and (\ref{N_factor})
\begin{equation}
J_{0}=e[1+2(t^{2}+\vect x^{2})+(x^{2})^{2}]\delta^{3}(\vect x)=e(1+t^{2})^{2}\delta^{3}(\vect x),\quad J_{i}=0,\label{N_J}
\end{equation}

with $E_{i}$ and $B_{i}$ still given by equation (\ref{fundamental}),
where we have omitted the prime in these notations. The region of
$N$ uncovered by the conformally flat coordinates corresponds to
the ordinary conformal boundary of the Minkowski spacetime. By the
inversion (\ref{inversion}), we can examine the uncovered region
(actually a ``compactified'' light cone). For the electric current,
we have
\begin{equation}
\tilde{J}_{0}(\tilde{x})=e(1+\tilde{t}^{2})^{2}\delta^{3}(\tilde{\boldmath x}),\quad\tilde{J}_{i}(\tilde{x})=0.\label{N_J_inverse}
\end{equation}
For the electromagnetic field, we have also equations (\ref{tilde_E})
and (\ref{tilde_B}). So we can see the breakdown of the Maxwell equation
at $\tilde{x}^{2}=0$, similar to the dS case. Unlike the dS case,
however, since $N$ has no (conformal) boundary, this breakdown cannot
be remedied by the sign reversion (\ref{reversion}) of $A_{\mu}$
and $J_{\mu}$ in certain regions of $N$. Although one can see equations
(\ref{N_J}) and (\ref{N_J_inverse}) as the only correct form of
a point-like source that satisfies the continuity equation, there
is no corresponding electromagnetic field that globally satisfies
the Maxwell equation. In other words, one cannot find fundamental
solutions to the Maxwell equation on $N$. This problem can be resolved
by cutting open $N$ along $\tilde{x}^{2}=0$ and sewing another $N$
(also cut open) onto it, which yields the double covering $2N$, as
expected.

\section{Concluding Remarks}

In Section 2, we review the ordinary conformal compactification of
the Minkowski, dS and Ads space times, where one can see that the
pseudo-sphere really can help him to understand the conformal compactification.
First, we get the intersection $\mathcal{M}$, which in fact is a
Minkowski spacetime, of the hyperplane$\mathcal{P_{\mathit{a}}}$
and the zero radius pseudo-sphere $\mathcal{N}$ in a (4+2)-dimensional
Minkowski spacetime. It's not difficult to see that the infinity points
of $\mathcal{M}$ lie on the hyperplane parallel to $\mathcal{P_{\mathit{a}}}$.
The compactification {[}$\mathcal{N}${]}, in which the $\mathit{O}$(2,4)/$\mathbb{Z}_{2}$
action can be well defined, is generated by adding those infinity
points to $\mathcal{M}$. That is the compactification of Minkowski
spacetime. For the dS and Ads case, we use a general hypersurface
(of antipodal symmetry) to intersect $\mathcal{N}$ to get some conformally
flat manifolds.

In Section 3, we map the solution to Maxwell equation
on Minkowski spacetime (\ref{fundamental}) to the dS spacetime to
get (\ref{dS_J}), from which one can find that there are really two
antipodal point charges. This is caused by the fact that the conformal
boundary $1+x^{2}$=0 separates the world line $\vect x=0$
into two parts. With the inversion (\ref{inversion}), one can find
that the field (\ref{tilde_E}) is discontinuous at $\tilde{x}^{2}=0$,
which shows the breakdown of the Maxwell equation on the compactification
of Minkowski spacetime. It is not hard to see that the breakdown is
caused by the modulus term in equation (\ref{fundamental}).
One should take care that in (\ref{dS_inverse_J}) there is not a
simple $\tilde{x}^{2}$ factor but a Jacobian factor when one writes
$\delta^{3}\left(\tilde{\vect x}/\tilde{x}^{2}\right)$ into $\delta^{3}$$\left(\tilde{\vect x}\right)$.
And the condition for magnetic charge is not different .

In Section 4, the discussion in Section 3 is illustrated with Penrose
diagrams. Reversing the sign of the electromagnetic field and current
(see (\ref{reversion})) in ``I'' to ``IV'' in figure (\ref{Penrose-InvdS}),
one can see the Maxwell equations can be well defined on the whole
dS spacetime. Equation (\ref{global}) is the global solution,
and one can see that there are two antipodal point charges. We also
show that the ordinary compactification, identifying ``I'' to ``IV''
with ``I$'$'' to ``IV$'$'', can not give a global defined electrodynamics.
This leads us to consider the doubled conformal
compactification, and we reveal that superposing
the familiar Penrose diagrams of dS and Minkowski space times (see
figure 2), and identifying the regions ``I'' to ``IV'' with ``I$'$''
to ``IV$'$'' as antipodal regions, respectively will give the doubled
compactification. Figures 3 and 4 show this process
clearly. And the doubled conformal compactification for the electrodynamics
with a magnetic charge is very similar to the electron, which we don't
analysis here.

Since our attention is solely paid to the classical
case here, one should further consider the conformal invariant quantum
field theory containing electrodynamics correspondingly \cite{M.BAKER AND K.JONSON,H.Osborn}.
And zero-mass systems \cite{McLennan1,MN,JV} can also be considered,
including the Lienard-Wiechert field of massless charges \cite{Francesco Azzurli and Kurt Lechner16}.
These problems should be left for future works. The CPT invariance
, and causality are also some interesting aspects
for our further studying.

\subsection*{Acknowledgments}

We would like to thank C.-G. Huang, X.-N. Wu, Z. Xu and B. Zhou for
helpful discussions. This work is partly supported by the National
Natural Science Foundation of China under Grant Nos. 11075206 and
11175245.

\end{document}